\documentclass[journal=jacsat,manuscript=article]{achemso}

\usepackage{amsmath}
\usepackage{amssymb}
\usepackage{xcolor}
\usepackage{epsf}
\usepackage{graphicx}
\usepackage{bm}

\author{Igor V. Bondarev}
\email{ibondarev@nccu.edu}
\affiliation[NCCU]
{Department of Mathematics and Physics, North Carolina Central University,\\ Durham, NC 27707, USA}

\title{Controlling Single-Photon Emission with Ultrathin Transdimensional Plasmonic Films}

\abbreviations{TD, CN, QED}

\begin{document}

\begin{abstract}
We study theoretically the properties of a two-level quantum dipole emitter near an ultrathin transdimensional plasmonic film. Our model system mimics a solid-state single-photon source device. Using realistic experimental parameters, we compute the spontaneous and stimulated emission intensity profiles as functions of the excitation frequency and film thickness, followed by the analysis of the second-order photon correlations to explore the photon antibunching effect. We show that ultrathin transdimensional plasmonic films can greatly improve photon antibunching with thickness reduction, which allows one to control quantum properties of light and make them more pronounced. Knowledge of these features is advantageous for solid-state single-photon source device engineering and overall for the development of the new integrated quantum photonics material platform based on the transdimensional plasmonic films.
\end{abstract}


\newpage

\subsubsection{Introduction}

Transdimensional (TD) quantum materials are atomically-thin films of precisely controlled thickness, films made of precisely controlled finite number of monolayers~\cite{Rivera,BondOMEX17,BoltShalACS19}. The term 'transdimensional' refers to the transitional range of thicknesses --- a regime that is neither three (3D) nor two (2D) dimensional but rather something in between, turning into 2D as thickness tends to zero, challenging to study what the 3D-to-2D continuous transition has to offer to improve material functionalities. Currently available due to the rapid progress in nanofabrication techniques~\cite{thingold,thinXenes,Shah17,Shah18,javierOptica19,GarciaAbajoACS19,TunableGold,NL22TiN}, such materials offer high tailorability of their electronic and optical properties not only by altering their chemical and/or electronic composition (stoichiometry, doping) but also by merely varying their thickness (number of monolayers)~\cite{Brongersma17,BondMRSC18,BondPRR20,Manjavacas22}. Materials like these are indispensable for studies of fundamental properties of the light-matter interaction as it evolves from a single 2D atomic layer to a larger number of layers approaching the 3D bulk material properties. With thickness of only a few atomic layers, ultrathin TD films of metals, doped semiconductors, or polar materials can support plasmon-, exciton-, magnon-, and phonon-polariton eigenmodes~\cite{BondPRR20,Manjavacas22,ManjavacasNatCom14,GarciaAbajoFD15,CampionePRB15,BondOMEX19,CNArr21PRAppl,CNArr21JAP,Miller17,BondVlad18,Thygesen18,Lius-PKim19,Geim20,
CommPhys-bond,Xu21,SnokeBond21,magnons,ZhelNatCom18,MariaNL19}. Plasmonic TD materials (ultrathin metallic films) offer controlled light confinement, large tailorability and dynamic tunability of their optical properties due to their thickness-dependent localized surface plasmon (SP) modes~\cite{BondPRR20,Manjavacas22,ManjavacasNatCom14,GarciaAbajoFD15,CampionePRB15,BondOMEX19,CNArr21PRAppl,CNArr21JAP}, which are distinctly different from those of conventional thin films commonly described by either purely 2D or by 3D material properties with boundary conditions imposed on their top and bottom interfaces~\cite{Ritchie,Economou,Dahl,Theis,Ando,Chaplik,Wang,Pitarke,Politano}. In such systems, the vertical quantum confinement enables a variety of new quantum phenomena, including the thickness-controlled plasma frequency red shift~\cite{BondOMEX17,NL22TiN}, the SP mode degeneracy lifting~\cite{BondPRR20,CampionePRB15}, a series of magneto-optical effects~\cite{BondMRSC18}, and even atomic transitions that are normally forbidden~\cite{Rivera,CNArr21PRAppl,CNArr21JAP}, to mention a few.

\begin{figure}[t]
\includegraphics[scale=0.85]{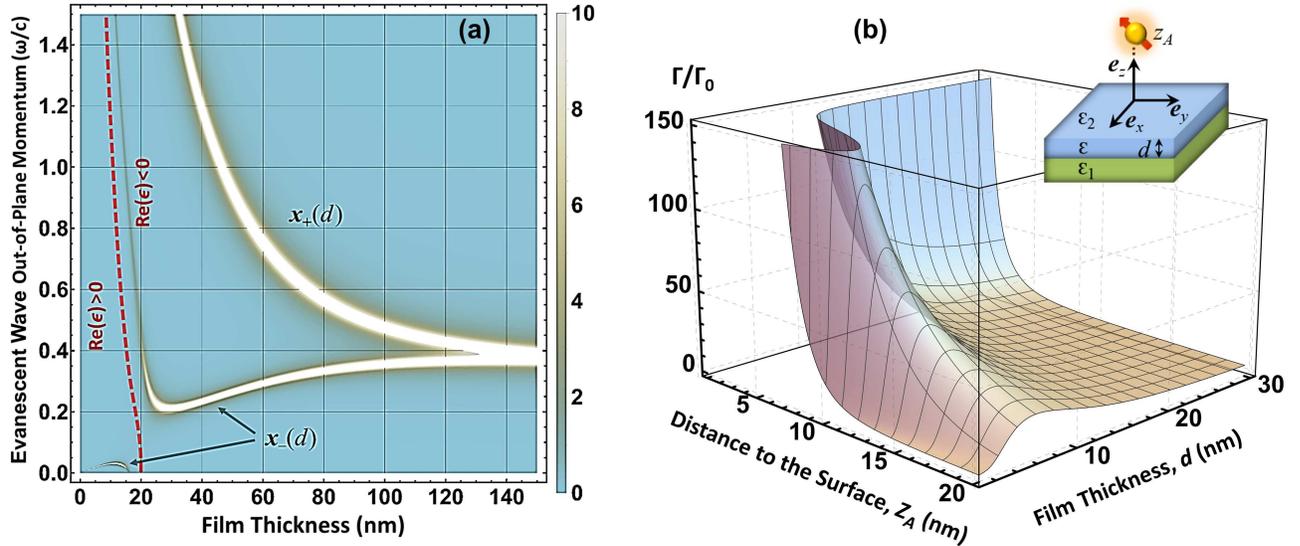}\\[0.5cm]
\caption{(a)~The plasma mode distribution derived from the $p$-evanescent wave reflection coefficient of the ultrathin TD plasmonic film as a function of the film thickness. (b)~The spontaneous emission rate (averaged over the three dipole orientations and normalized by the spontaneous emission rate in vacuum) for the DE coupled to the plasma modes in (a), as a function of the film thickness $d$ and the DE position $z_A$ near the TD film surface where the evanescent contribution to the spontaneous emission is dominant. Inset shows the geometry of the problem. See Ref.~\cite{BondPRR20} for details.}
\label{fig1}
\end{figure}

Previously, we have implemented the confinement-induced nonlocal Drude response model based on the Keldysh-Rytova (KR) pairwise electron interaction potential~\cite{BondOMEX17} to study theoretically the electronic properties of ultrathin TD plasmonic films~\cite{BondMRSC18,BondPRR20} and demonstrate their major manifestations experimentally~\cite{NL22TiN,Lavrinenko19}. The KR interaction potential takes into account the vertical electron confinement~\cite{KRK,KRR}, which makes it much stronger than the electron-electron Coulomb potential~\cite{KRK}, offering also the film thickness as a parameter to control the electronic properties of the TD plasmonic films. Here, we use the KR model to further explore the capabilities of the TD plasmonic films, focusing on the light-scattering properties of a quantum dipole emitter (DE) placed at a distance near the film surface; see the inset in Fig.~\ref{fig1}~(b). The model system we study mimics a quasi-2D solid-state single-photon source device,~\cite{Aharon22} for which our goal is to show the advantages of using the ultrathin TD plasmonic films. So far room-temperature solid-state quantum emitters have been observed in wide-bandgap semiconductors such as diamond~\cite{Aharon14} and silicon carbide~\cite{Castelletto14}, nanocrystalline quantum dots~\cite{Mangum14,Pisanello13,Michler00}, and in carbon nanotubes~\cite{Htoon15}. Single-photon emission from localized color-center defects in 2D materials has been reported both at cryogenic~\cite{Chakraborty15,He15,Kopersk15,Srivastava15} and at room temperatures~\cite{Aharon16,Aharon20}. An aspiring goal, however, is to use these materials in the quantum regime where the photon emission is of robust and controllable single-photon character, to enable applications in quantum information processing~\cite{Aharon22,Vuchkovich09}. We show that the ultrathin TD plasmonic films have this outstanding potential due to their unique tailorability by means of their thickness adjustment.

Using the nanodiamond nitrogen-vacancy (NV) center~\cite{SimeonBogdanov,NVtransition} near the TiN film surface as a prototype coupled 'DE-TD film' system~\cite{BondPRR20}, we compute the spontaneous and stimulated emission intensity profiles as functions of the excitation frequency and film thickness, followed by the intensity correlation function analysis to explore the photon antibunching effect. We show that the film thickness can be used to tune from weak to strong the evanescent coupling of the DE to the plasma modes of the ultrathin TD films. The strong evanescent coupling hybridizes the transition energy levels of the DE whereby its stimulated emission is quenched. Increased resonance optical pumping reduces this quenching effect. The overall film thickness effect on the intensity correlation function is to increase its positive slope with thickness reduction and thus to improve the photon antibunching and related (nonclassical) sub-Poissonian photon counting statistics~\cite{Cook81}. Knowledge of these features is advantageous for solid-state single-photon source device engineering~\cite{Aharon22} and overall for the development of the new integrated quantum photonics material platform~\cite{MaterPlatf} based on the ultrathin TD materials for quantum information processing applications.

\subsubsection{a. Spontaneous Emission Spectrum}

Figure~\ref{fig1}~(a) shows the earlier reported plasma mode distribution derived from the $p$-evanescent wave reflection coefficient of the ultrathin TD plasmonic slab as a function of the thickness $d$ of the slab~\cite{BondPRR20}. The associated spontaneous emission rate (averaged over the three dipole orientations and normalized by the spontaneous emission rate in vacuum) is shown in Fig.~\ref{fig1}~(b) as a function of $d$ and the DE position $z_A$ near the surface (see inset) where the evanescent contribution to the spontaneous emission is dominant. The DE transition frequency is taken to be $1.81$~eV consistent with what was reported recently from the density functional theory simulations for the zero-phonon line transition of the NV-center in diamond~\cite{NVtransition}, with all other parameters being essentially representative of the ultrathin TiN film as reported earlier~\cite{BondPRR20}. Comparing the (a) and (b) panels, it can be clearly seen that the fast increase followed by the drop-off of the DE spontaneous emission rate at $d\!<\!10$~nm originates from the DE coupling to the split-off plasma mode of the lowest out-of-plane momentum which is present in there for $10\gtrsim d\gtrsim 5$~nm. The presence itself of this split-off mode is a remarkable feature of the confinement-induced nonlocality captured by the KR model we use~\cite{BondPRR20,NL22TiN}.

To see how the DE coupling to the plasma modes in Fig.~\ref{fig1}~(a) affects its spontaneous emission intensity profile, we use the analytical approach previously developed by one of us for an analogous problem of the DE coupled to a resonance of the local density of photonic states (LDOS) close to the nanotube surface~\cite{Bond06PRB}. In our notations here, the spectral lineshape profile of the spontaneous emission intensity is given by
\begin{equation}
I(d,z_A,\omega)=\frac{\Gamma(d,z_A)}{2\pi}\frac{(\omega-\omega_{A})^{2}+\delta\omega_{r}^{2}} {\left[(\omega-\omega_{A})^{2}-X^{2}/4\right]^{2}+\delta\omega_{r}^{2}(\omega-\omega_{A})^{2}}\,,
\label{Ixfin}
\end{equation}
\[
X(d,z_A)=\sqrt{2\delta\omega_r\Gamma(d,z_A)}\,.
\]
Here, the DE transition frequency $\omega_{A}$ includes the red shift due to the DE interaction with the surface~\cite{Bond05PRB} and is assumed to be dimensionless (divided by $\Gamma_0$, the vacuum spontaneous emission rate~\cite{VogelWelschQM}), and so too are all quantities in the equation including the half-width-at-half maximum $\delta\omega_{r}$ of the LDOS resonance. The thickness-dependent spontaneous emission rate $\Gamma(d,z_A)\!=\Gamma(d,z_A,\omega_A)/\Gamma_0$ is the one presented in Fig.~\ref{fig1}~(b). The parameter $X$ is the Rabi-splitting to represent the DE transition level hybridization due to the coupling to the medium-assisted excitations of the material subsystem (plasma modes of the TD film herein). The coupling is termed weak if $(X/\delta\omega_{r})^2\!\ll\!1$ and strong if $(X/\delta\omega_{r})^2\!\gg\!1$, in which cases Eq.~(\ref{Ixfin}) generates the single-peaked $\omega_A$-centered Lorentzian profile and the double-picked profile of two symmetrically split peaks at $\omega_A\pm X/2$, respectively~\cite{Bond06PRB}.

\subsubsection{b. Stimulated Emission Spectrum}

In resonance-fluorescence experiments, the DE is irradiated by laser light tuned to the DE transition frequency, so that the DE can be excited into an upper quantum state. The excitation is followed by two competing processes. They are the (coherent) driving of the DE transition by the laser field, whereby stimulated (coherent) photon emission and absorption of a laser photon take turns, and spontaneous (incoherent) emission of the DE excited state. An interplay of the two, of which one is strongly $d$-dependent in our case (Fig.~\ref{fig1}), is expected to give rise to new quantum-statistical features of the scattered radiation.

It was first recognized by R.Feynmann et al.~\cite{FeynmanOptBloch} that the resonance-fluorescence process can be described by optical Bloch equations, a generalization of the Bloch equations of the magnetic resonance~\cite{Blum}, whereby the DE stimulated emission spectrum under resonance excitation can be shown to take the form as follows~\cite{VogelWelschQM}
\begin{equation}
S(\omega,\Gamma_f)=\frac{1}{\pi}\,\mbox{Re}\Big\{\tilde{S}_{12}\Big[i\big(\omega-\omega_L\big)+\frac{\Gamma_f}{2}\Big]\Big\}\,.
\label{S}
\end{equation}
Here, $\omega_L$ is the laser field frequency, $\Gamma_f$ is the spectral apparatus passband width, and
\begin{equation}
\tilde{S}_{12}(s)=\frac{\sigma_{22}(\infty)}{s+\Gamma_2}\Big\{1+\frac{\Gamma_1}{2s}\Big[1-\frac{\Omega/\Gamma_1}{\Omega/(s+\Gamma_1)+(s+\Gamma_2)/\Omega}\Big]\Big\}
\label{S12}
\end{equation}
is the Laplace transform of the spectral distribution function under the resonance ($\omega_A\!=\omega_L$) single-photon transition $|1\rangle|n+1\rangle\!\leftrightarrow|2\rangle|n\rangle$ ($n$ is the photon occupation number) between the lower $|1\rangle$ and upper $|2\rangle$ DE states. In this equation, $\Omega\!=|\textbf{d}_{21}\!\cdot\textbf{E}_L|/2\hbar$ is the Rabi frequency to represent the DE transition dipole coupling strength to an external laser field, $\Gamma_1$ and $\Gamma_2$ are the longitudinal and transverse relaxation rates, respectively, analogous to those of the Bloch magnetic resonance theory~\cite{Blum}. In our case here, we have $\Gamma_1\!=2\Gamma_2\!=\Gamma(d,z_A)$ since major relaxation comes due to spontaneous emission from the upper DE state~\cite{VogelWelschQM}. The prefactor
\begin{equation}
\sigma_{22}(\infty)=\frac{1}{2}\frac{\Omega^2}{\Gamma_1\Gamma_2+\Omega^2}
\label{sigma22}
\end{equation}
is the time-dependent upper-state occupation probability taken at $t\!=\!\infty$. All quantities in Eqs.~(\ref{S})--(\ref{sigma22}) are assumed to be dimensionless as per convention introduced above for Eq.~(\ref{Ixfin}).

Equations~(\ref{S})--(\ref{sigma22}) allow for two distinct limiting cases~\cite{VogelWelschQM}. They are (1)~the weak-driving-field limit, where $\Omega\!\ll\!\Gamma_1,\Gamma_2$ to result in the $\omega_L$-centered ($\omega_L\!=\omega_A$) single-peak Rayleigh scattering spectral profile, and (2)~the high-driving-field limit, where $\Omega\!\gg\!\Gamma_1,\Gamma_2$ and the spectrum is reconstructed due to the "dressed" state formation to yield the Mollow triplet profile consisting of the $\omega_L$-centered main peak and the two symmetric satellites at $\omega_L\pm\Omega$.

\subsubsection{c. Intensity Correlation and Photon Antibunching}

Having emitted a photon, the DE undergoes a quantum jump from the upper to the lower (ground) quantum state, whereby it can no longer emit a second photon. Only after performing a laser-induced transition back to the upper quantum state, the DE is ready to emit a second photon. The probability of emitting the first photon at an instant of time $t$ is proportional to the upper-state photon occupation probability $\sigma_{22}(t)$, which also controls the intensity $S$ of the scattered light as per Eqs.~(\ref{S})--(\ref{sigma22}), and which can be found from the optical Bloch equations to have the form as follows~\cite{VogelWelschQM}
\begin{equation}
\sigma_{22}(t)=\sigma_{22}(\infty)\Big(1+\frac{\lambda_2\,e^{\lambda_1t}}{\lambda_1-\lambda_2}+\frac{\lambda_1\,e^{\lambda_2t}}{\lambda_2-\lambda_1}\Big)\,.
\label{sigma22t}
\end{equation}
Here
\begin{equation}
\lambda_{1,2}=-\frac{1}{2}\big(\Gamma_1+\Gamma_2\big)\pm\sqrt{\frac{1}{4}\big(\Gamma_1-\Gamma_2\big)^2-\Omega^2}\,,
\label{lambda12}
\end{equation}
and the DE is assumed to be resonantly ($\omega_L\!=\omega_A$) excited out of its initial lower (ground) quantum state. Similarly, the probability of emitting a second photon at a later instant of time $t+\tau$ is proportional to the DE upper-state occupation probability $\sigma_{22}(t+\tau)$ under the condition that at time $t$ the DE is in its lower (ground) quantum state. Hence, the probability of emitting a first photon at time $t$ and a second photon at time $t+\tau$ is proportional to $\sigma_{22}(t)\,\sigma_{22}(t+\tau)$. Under the steady-state observation conditions, this turns into the two-photon scattered-intensity correlation function $\sigma_{22}(\infty)\,\sigma_{22}(t+\tau)|_{t\rightarrow\infty}$, a quantity directly related to the normalized second-order autocorrelation function
\begin{equation}
g^{(2)}(\tau)=\frac{\sigma_{22}(|\tau|)}{\sigma_{22}(\infty)}=1+\frac{\lambda_2\,e^{\lambda_1|\tau|}}{\lambda_1-\lambda_2}+\frac{\lambda_1\,e^{\lambda_2|\tau|}}{\lambda_2-\lambda_1}
=1-(1+a)\,e^{\lambda_1|\tau|}+a\,e^{\lambda_2|\tau|}
\label{gamma22}
\end{equation}
with $a\!=\!\lambda_1/(\lambda_2-\lambda_1)$, commonly used in single-photon source characterization experiments~\cite{Aharon16}.

The function $g^{(2)}(\tau)$ can be seen to have a positive initial slope, which reflects the quantum optics effect of photon antibunching. Clearly, the greater this slope is, the more pronounced quantum features of light are. In addition, it can be shown that since $g^{(2)}(\tau)<1$, the scattered fluorescence light gives rise to a (nonclassical) sub-Poissonian photocounting statistics~\cite{Cook81}. Both photon antibunching nature and the sub-Poissonian statistics of the resonance fluorescence from a two-level DE system may be regarded as proofs of the quantum nature of light which cannot be understood and explained in terms of classical optics.

\begin{figure}[t]
\includegraphics[scale=0.85]{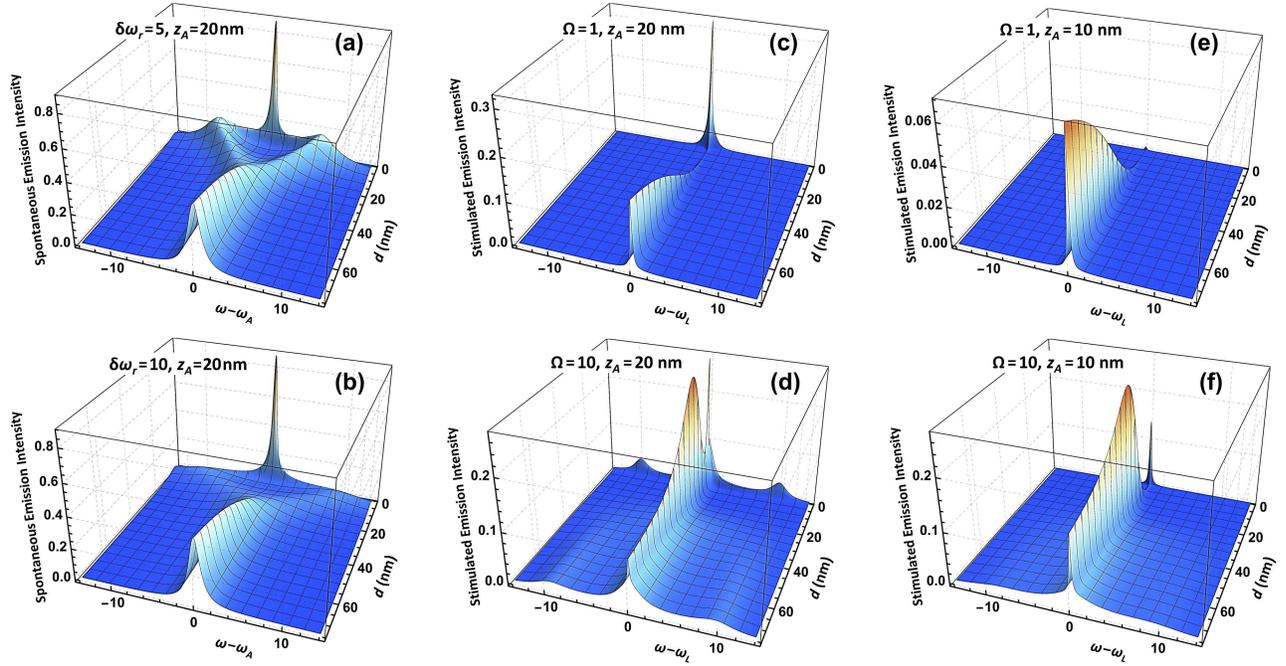}\\[0.5cm]
\caption{(a),(b) Spontaneous emission spectra calculated from Eq.~(\ref{Ixfin}) with two representative $\delta\omega_r$ and $\Gamma(d,z_A)$ shown in Fig.~\ref{fig1}~(b) for the DE positioned at $z_A\!=20$~nm and coupled to the TD film plasma modes shown in Fig.~\ref{fig1}~(a). (c),(d) Stimulated emission spectra calculated from Eqs.~(\ref{S})--(\ref{sigma22}) with two representative $\Omega$ and $\Gamma(d,z_A)$ shown in Fig.~\ref{fig1}~(b) for the DE positioned at $z_A\!=20$~nm and coupled to the TD film plasma modes shown in Fig.~\ref{fig1}~(a). (e),(f) Same as in (c),(d) for the DE positioned at $z_A\!=10$~nm.}
\label{fig2}
\end{figure}

\subsubsection{d. Discussion of Numerical Results}

Figure~\ref{fig2}~(a) presents the spontaneous emission spectrum calculated from Eq.~(\ref{Ixfin}) with $\delta\omega_r\!=5$ (taken as a model parameter) and $\Gamma(d,z_A)$ shown in Fig.~\ref{fig1}~(b) for the DE positioned at $z_A\!=20$~nm and coupled to the TD film plasma modes shown in Fig.~\ref{fig1}~(a). Figure~\ref{fig2}~(b) presents the same for $\delta\omega_r\!=10$ to compare. The most interesting feature there is that as $d$ decreases, in the domain $20\gtrsim d\gtrsim5$~nm, the DE-plasmon coupling strengthens to result in the fast spontaneous emission rate increase, as can be seen in Fig.~\ref{fig1}~(b), whereby the coupled system is driven in the strong coupling regime $(X/\delta\omega_{r})^2\!\gg\!1$ where the new set of (hybridized) eigen levels forms~\cite{Bond06PRB}, $\omega_A\!\pm X/2$. This turns the single-peaked $\omega_A$-centered Lorentzian spontaneous emission profile into the double-picked profile of the two symmetrically split resonances at $\omega_A\!\pm X/2$. Increased $\delta\omega_r\!=10$ does not eliminate but just smoothes out this effect. In this domain of film thicknesses, $\omega_A$ is no longer the DE eigen transition frequency, and so the intensity of the stimulated emission Rayleigh (same frequency) scattering profile drops down, as can be seen in Fig.~\ref{fig2}~(c) for the same $z_A$ and $\Omega=1$, the DE radiative coupling to the external laser field. Instead, it can be shown that the plasmon enhanced Raman (shifted frequency) scattering is very efficient in this regime~\cite{Bond15OE}. Figure~\ref{fig2}~(d) shows that the increased DE-laser field coupling, $\Omega=10$, can compete with the DE-plasmon coupling to restore the stimulated Rayleigh scattering intensity by driving the system in the strong radiative coupling regime with "dressed" states formed where the representative Mollow triplet spectral profile appears. Figure~\ref{fig2}~(e), where $z_A$ is decreased down to $10$~nm, demonstrates that bringing the DE closer to the film surface does drop down the intensity of the stimulated emission Rayleigh scattering profile for $\Omega=1$. The obvious reason is that the DE-plasmon coupling increases as well (not shown) so that even increased $\Omega=10$ can barely restore it completely as can be seen from Fig.~\ref{fig2}~(f). Finally, decreasing $d$ below $5$~nm restores the sharp central peaks for both spontaneous and stimulated emission profiles in Fig.~\ref{fig2} as only one plasma mode is found in this domain in Fig.~\ref{fig1}~(a), which dies with $d$ decreasing to reduce $\Gamma(d,z_A)$ as can be seen in Fig.~\ref{fig1}~(b). This brings our system back in the weak DE-plasmon coupling regime, $(X/\delta\omega_{r})^2\!\ll\!1$, followed by making it transparent in the $d\!\rightarrow0$ limit.

\begin{figure}[t]
\includegraphics[scale=0.85]{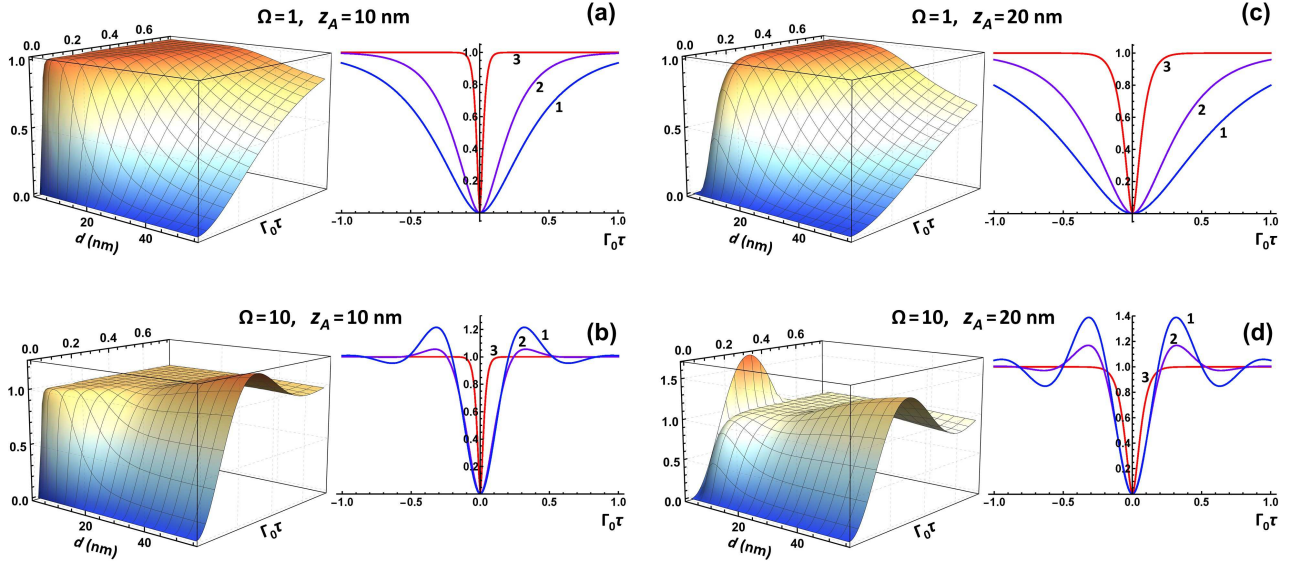}\\[0.5cm]
\caption{(a),(b) Normalized second-order autocorrelation function $g^{(2)}(\tau)$ calculated from Eq.~(\ref{gamma22}) with two representative $\Omega$ and $\Gamma(d,z_A)$ shown in Fig.~\ref{fig1}~(b) for the DE positioned at $z_A\!=10$~nm and coupled to the TD film plasma modes shown in Fig.~\ref{fig1}~(a). ~(c),(d) Same as in (a),(b) for the DE positioned at $z_A\!=20$~nm. Left panels show the 3D graphs of the functions $g^{(2)}(\tau\!>\!0)$; right panels are their crosscuts by planes of constant $d=50$~nm (line~1), $d=30$~nm (line~2), and $d=10$~nm (line~3).}
\label{fig3}
\end{figure}

Figure~\ref{fig3} presents our calculations of the normalized second-order autocorrelation function $g^{(2)}(\tau)$ of Eq.~(\ref{gamma22}) for two representative $\Omega=1$ and $10$ with $\Gamma(d,z_A)$ shown in Fig.~\ref{fig1}~(b), for the DE placed at $z_A\!=10$~nm and $20$~nm and coupled to the TD film plasma modes shown in Fig.~\ref{fig1}~(a). In all four graphs, the left panels show $g^{(2)}(\tau\!>\!0)$ as functions of $d$ and $\Gamma_0\tau$, while the right panels are their crosscuts by planes of constant $d=50$~nm (line~1), $d=30$~nm (line~2), and $d=10$~nm (line~3). There are two regimes that can be clearly seen there. They are the weak-driving-field regime and the strong-driving field regime. The former occurs for $\Omega=1$ in the entire domain of $d$ presented. Here $g^{(2)}(\tau)$ (and the two-photon scattered-intensity correlation function, accordingly) is seen to grow up monotonically from the value $g^{(2)}(0)=0$ to approach the stationary value of $g^{(2)}(\infty)=1$, with characteristic time scale controlled by the DE spontaneous relaxation times $\Gamma_1=\Gamma(d,z_A)$ and $\Gamma_2=\Gamma(d,z_A)/2$. The latter takes place for $\Omega=10$, where the function $g^{(2)}(\tau)$ oscillates at Rabi frequency in the domain of moderate $d$, with Rabi oscillations being damped due to the DE spontaneous relaxation processes. With decrease of $d$, however, in both cases the sharp photon antibunching can be seen in the ultrathin film thickness regime. The positive initial slope of the function $g^{(2)}(\tau)$ increases with the film thickness decrease to make the quantum single-photon features of light more pronounced for ultrathin TD films.

\subsubsection{Conclusion}

In this work we use the parameters of the nanodiamond NV center near the TiN film surface to compute the spontaneous and stimulated emission intensity profiles as functions of the excitation frequency and film thickness, followed by the intensity correlation function analysis to explore the photon antibunching effect for the quantum DE coupled the ultrathin TD plasmonic film. We show that the film thickness can be used to tune the evanescent coupling of the DE to the plasma modes of the film. The strong evanescent coupling hybridizes the transition energy levels of the DE, whereby its stimulated emission Rayleigh scattering profile is quenched. Increased resonance optical pumping drives the system in the strong radiative coupling regime, which competes with the DE-plasmon coupling, to restore the stimulated Rayleigh scattering intensity. The overall film thickness effect on the second-order photon autocorrelation function is to increase its positive slope with thickness reduction and thus to improve the photon antibunching properties and related (nonclassical) sub-Poissonian photon counting statistics. Knowledge of these features is advantageous for solid-state single-photon source device engineering and overall for the development of the new integrated quantum photonics material platform based on the ultrathin TD materials for quantum information processing applications.


\begin{acknowledgement}
This research is supported by the U.S. National Science Foundation under Condensed Matter Theory Program Award No.~DMR-1830874 (I.V.B.). I.V.B. acknowledges hospitality of the Kavli Institute for Theoretical Physics (KITP), UC Santa Barbara, where this work was completed during his invited visit as a KITP Fellow 2022--23 (the research program supported by the U.S. National Science Foundation under Grant No. PHY-1748958).
\end{acknowledgement}





\end{document}